\RequirePackage{fix-cm}
\documentclass[smallextended]{svjour3}       
\newcommand{\EDIT}{}		
\usepackage{ulem}
\smartqed  
\usepackage{graphicx}
\begin{document}
\title{INVITED PAPER\thanks{The UBC authors acknowledge support from the National Science and Engineering Research Council and the British Columbia Innovation Council.}
}
\subtitle{Design and modeling of a transistor vertical-cavity surface-emitting laser}


\author{Wei Shi \and Behnam Faraji \and  Mark Greenberg \and
        Jesper Berggren \and Yu Xiang \and Mattias Hammar \and
        Michel Lestrade \and Zhi-Qiang Li \and Z. M. Simon Li \and Lukas Chrostowski         
}
%




\institute{Wei Shi, Behnam Fariji, Mark Greenberg, and Lukas Chrostowski* \at
              Department of Electrical and Computer Engineering, The University of British Columbia, Vancouver, BC, Canada
              ~* \email{lukasc@ece.ubc.ca}           
           \and
           Jesper Berggren, Yu Xiang, and Mattias Hammar \at
              School of Information and Communication Technology (ICT), The Royal Institute of Technology (KTH), 16440 Kista, Sweden
            \and
            Michel Lestrade, Zhi-qiang Li, and Z. M. Simon Li  
           Crosslight Software Inc., 121-3989 Henning Dr. Burnaby, BC , V5C 6P8 Canada
}

\date{Received: Oct. 4, 2010 / Accepted: Jan. 25, 2011}

\maketitle

\begin{abstract}
A multiple quantum well (MQW) transistor vertical-cavity surface-emitting laser (T-VCSEL) is designed and numerically modeled. The important physical models and parameters are discussed and 
validated by modeling a conventional VCSEL and comparing the results with the experiment. The quantum capture/escape process is simulated using the quantum-trap model and shows a significant effect on the electrical output of the T-VCSEL. The parameters extracted from the numerical simulation are imported into the analytic modeling to predict the frequency response and simulate the large-signal modulation up to 40 Gbps.
\end{abstract}

\section{Introduction}
While most semiconductor lasers are diode structures, transistor lasers \cite{TL} are attracting much attention for their potential for high-frequency operation \cite{Microwave} and unique carrier dynamics \cite{CB}. Different from monolithically integrating a transistor and a laser diode, either in-plane \cite{MonoInteg} or vertically \cite{VertiInteg}, where the transistor performs as the electrical driver of the laser, transistor lasers use the base recombination to provide stimulated emission. Early works on transistor lasers include a conceptual proposal of quantum well (QW) transistor lasers \cite {QW_LT_old} and experimental demonstration of a laser transistor \cite{LaserTransistor} that could function as a laser and a transistor in separate states. The recently developed QW transistor laser \cite{TL} has two independent control signals and can simultaneously output an electrical signal and an optical signal. This makes possible a simpler method of implementing feedback operation \cite{TVCSEL_feedback}, a unique voltage-controlled mode \cite{TXVCSEL}, and a new optoelectronic integration scheme. Compared with edge-emitting lasers, vertical-cavity surface-emitting lasers (VCSELs) have many attractive features \cite {VCSEL_advances} such as low power consumption, large-scale 2D array fabrication, on-wafer testing, single longitudinal mode operation, and a narrow circular beam. Integrating a heterojunction bipolar transistor (HBT) structure into a vertical cavity, transistor VCSELs (T-VCSELs) \cite{TXVCSEL} could combine the optoelectronic properties of the transistor laser with the advantages of VCSELs. 

In this work, we design and model a multiple QW (MQW) T-VCSEL self-consistently. The important parameters are calibrated by matching the simulation to the experiment of a conventional VCSEL. We link the numerical model with our previously developed analytic model \cite{Analytical}, using the quantum-trap model, to predict its frequency response and large-signal-modulation performance under different transistor bias configurations.
\section{Design}
\begin{figure}
\centering
\includegraphics[width=3.6in]{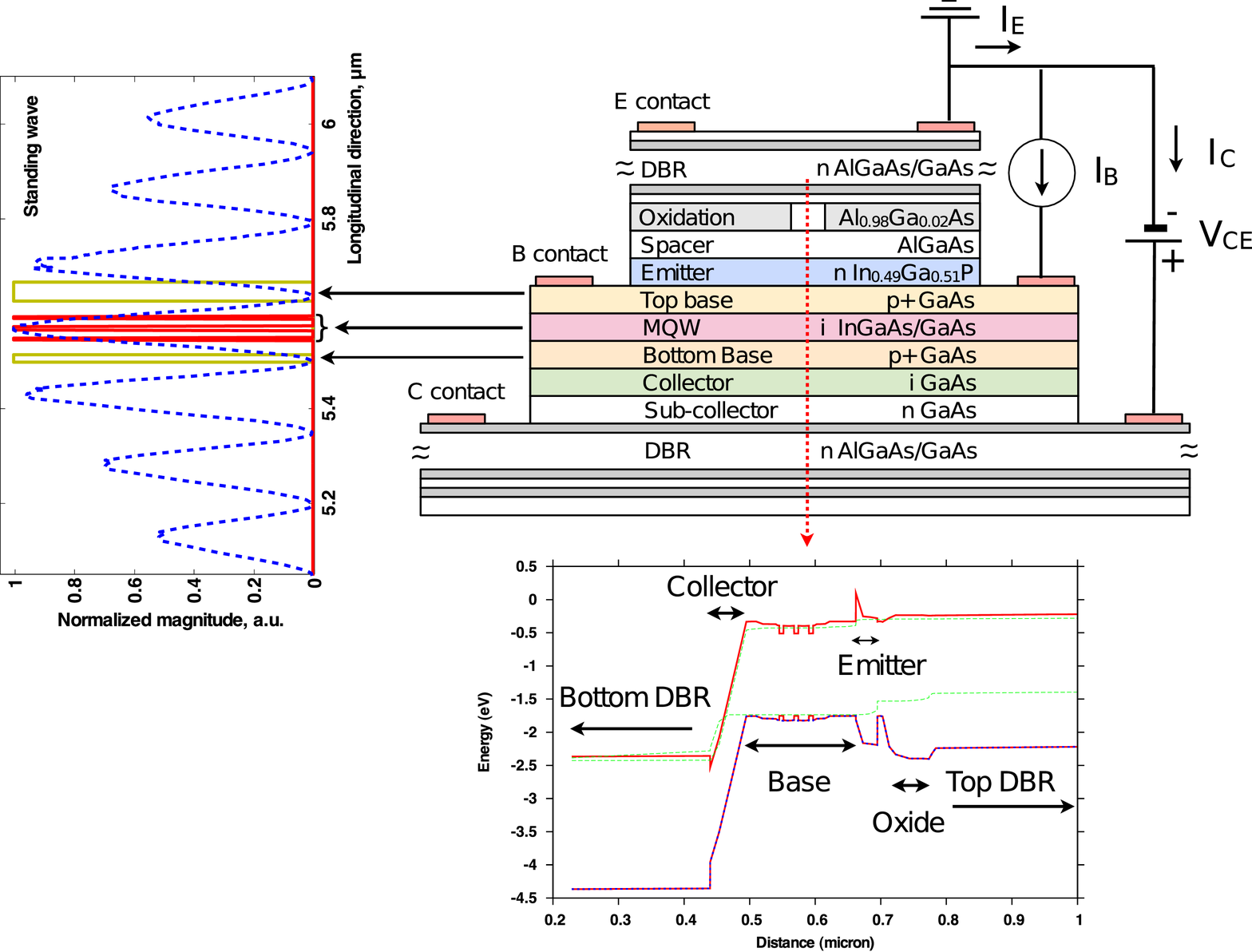}
\caption{Structure of the T-VCSEL with the DC bias configuration labeled. The insets show the standing wave (left) in the cavity with the positions of the QWs and heavily doped layers labeled and the typical band diagram (bottom) above threshold at the device center.} 
\label{structure}
\end{figure}

The T-VCSEL, as shown in Fig. \ref{structure}, has an Npn In$_{0.49}$Ga$_{0.51}$P/GaAs HBT structure. The bottom and top distributed Bragg reflectors (DBRs) consist of 36 pairs and 24 pairs of Al$_{0.85}$Ga$_{0.15}$As/GaAs, respectively. The base region plays a critical role in determinating the electrical and optical performance of a transistor laser \cite{TL_OE}. In this design, we use an asymmetric base doping profile where the whole base region is composed of (from bottom to top) a 15 nm heavily doped ($1\times10^{19}~cm^{-3}$) layer, a 30 nm doping grading layer, three intrinsic In$_{0.17}$Ga$_{0.83}$As/GaAs QWs, another 30 nm doping grading layer, and a 40 nm heavily doped ($1\times10^{19}~cm^{-3}$) base-contact layer. The heavily doped layers are aligned with the nodes of the longitudinal standing wave (as shown in the inset of Fig. \ref{structure}) in the vertical cavity to reduce the optical absorption. A $6 ~\mu m$-diameter oxide aperture is used for electrical and optical confinement.
\section{Numerical modeling}
\begin{figure}
\centering
\includegraphics[width=4.in]{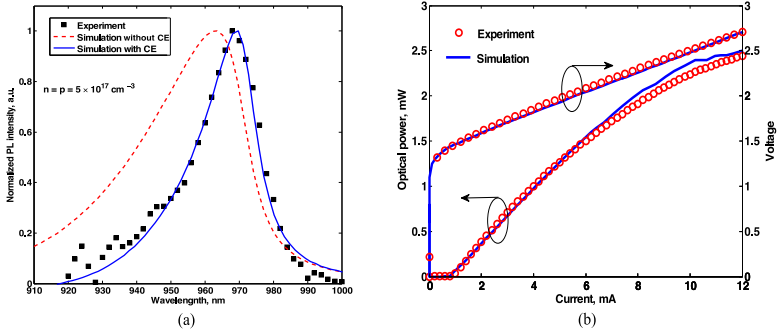}
\caption{Simulation vs. measurement of an InGaAs MQW VCSEL: (a) PL spectrum, with and without Coulomb enhancement (CE) in the simulation; (b) \emph{L-I-V} curves.}
\label{LIV_VCSEL}
\end{figure}

To investigate the physics and performance of the T-VCSEL, we have used PICS3D, an advanced numerical simulation software \cite{Crosslight}, that solves the electrical and optical models numerically and self-consistently. The carrier transport is described based on the classic drift-diffusion model \cite{Semiconductor_physics}, with the thermionic-emission model \cite{Semiconductor_physics} used at the heterojunctions. Lateral optical modes are calculated by the effective-index method \cite{EEM}. In the strained QWs, the conduction bands are assumed to be parabolic, and the valence bands are calculated by the 6$\times$6 kp method for the valance-band mixing \cite{kp_method}. The important parameters for QWs are carefully calibrated to match the measured photoluminescence (PL): \EDIT{while the reported values of the band-edge offset for strained InGaAs/GaAs QWs have a big spread \cite{III-V_Band}}, it is assumed to be $\bigtriangleup E_c: \bigtriangleup E_v: 0.8 : 0.2$ \cite{Band_offset_InGaAs}; a scattering time of 85 fs is used in Lorentz broadening; bandgap renormalization is considered as a function of the local carrier density of electrons (n) and holes (p): $\bigtriangleup E_g = 3\times 10^{-10}~eV/m \cdot (n/2+p/2)^{1/3}$. With Coulomb enhancement \cite{ManyBody} included in calculating the optical gain and spontaneous emission, good agreement  is achieved for the PL spectrum, as shown in Fig. \ref{LIV_VCSEL} (a). The optical loss due to the free-carrier and intervalence-band absorption is calculated by $\alpha=k_n\cdot n +k_p\cdot p$ with $k_n=3\times10^{-18}~cm^{2}$ and $k_p=6\times10^{-18}~cm^{2}$ for the passive layers \cite{Model_GaAs_detector} and $k_n=6\times10^{-18}~cm^{2}$ and $k_p=14\times10^{-18}~cm^{2}$ for the QWs \cite{Model_980_pump}. The Auger recombination is calculated by $R_{aug}=C_p\cdot n \cdot p^2$ by assuming that the CHHS Auger process dominates \cite{Auger_InP}. The Auger coefficient $C_p$ is assumed to be $6.5\times10^{-30}~cm^6\cdot s^{-1}$ for the passive layers \cite{Auger_GaAs}  and $1\times10^{-29}~cm^6\cdot s^{-1}$ for the QWs \cite{Model_980_VCSEL}. To verify the models and determine the material parameters, we have fabricated and simulated a conventional oxide-confined VCSEL that has the same QW structure as the designed T-VCSEL, taking self-heating into consideration. The power-law model \cite{Model_heterostructure} is used for AlGaAs thermal conductivity. The thermal conductivity ($\kappa$) of the AlGaAs/GaAs DBRs as a function of temperature ($T$) is given by $C^\kappa_{sct}\cdot (\kappa_{AlGaAs}(T)/2+\kappa_{GaAs}(T)/2)$ where $C^\kappa_{sct}$ is used to describe the effect of phonon scattering \cite{Thermal_DBR} due to the DBR interfaces and is tuned to be 0.5 for the lateral direction and 0.4 for the vertical direction. The carrier mobility is also affected by the phonon scattering and 75-percent reduction of the average mobility of AlGaAs and GaAs for the DBRs gives a good agreement with measured LIV curves, as demonstrated in Fig. \ref{LIV_VCSEL} (b). Thermal effect is not considered in the T-VCSEL modeling.
\begin{figure}
\centering
\includegraphics[width=2.2in]{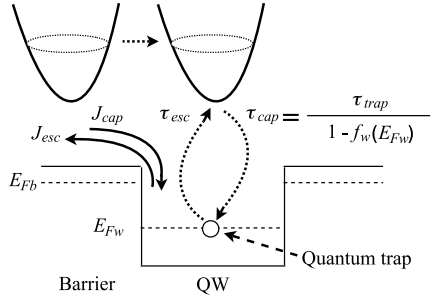}
\caption{Illustration of the QW capture/escape process and the quantum-trap model.} \label{trap}
\end{figure}

The QW capture/escape process has a significant influence on the frequency response of QW diode lasers \cite{Transiport_review} as well as of transistor lasers \cite{Analytical}. In this work, this process is described by a quantum-trap model in which the QWs are treated as carrier traps with trapping rates determined by phonon scattering theory \cite{q_transport}. The capture current ($J_{cap}$) and escape current ($J_{esc}$) of electrons (the minority carriers in the base) are given by 
\begin{eqnarray}
\label{eq:capture}
J_{cap} (E_{F_b},E_{F_w}) &=& \frac{1}{\tau_{cap}}\cdot d_w\cdot n_b(E_{F_b}) \nonumber \\
 &=&\frac{1}{\tau_{trap}}\cdot d_w\cdot n_b(E_{F_b})\cdot [1 -f_w(E_{F_w})]
\end{eqnarray}
\begin{equation}
\label{eq:escape}
J_{esc} (E_{F_w})=\frac{1}{\tau_{esc}}\cdot d_w\cdot n_w(E_{F_w})
\end{equation}
%
%
where $\tau_{cap}$, $\tau_{trap}$, and $\tau_{esc}$ are capture time, trapping time, and escape time, respectively; $d_w$ is the QW thickness; $n$, $E_F$, and $f$ represent carrier density, quasi-Fermi level, and occupancy, respectively, with the subscripts $w$ indicating ``well'' and $b$ ``barrier". From Eq. \ref{eq:capture} we see that $\tau_{cap}$ can be determined from $\tau_{trap}$ based on the Fermi-Dirac distribution. In addition, $\tau_{esc}$ can be calculated by $\tau_{cap}$ in the quasi-equilibrium condition in which $E_{F_b}$ is assumed to be equal to $E_{F_w}$ and hence $J_{esc}=J_{cap}$. Therefore, from Eq. \ref{eq:capture} and \ref{eq:escape} we can find $\tau_{esc}$:
\begin{equation}
\label{eq:tau_esc}
\frac{1}{\tau_{esc}} = \frac{1}{\tau_{trap}}\cdot \frac{n_b(E_{F_w})}{n_w(E_{F_w})} \cdot [1-f_w(E_{F_w})]
\end{equation}
\section{DC performance}
\begin{figure}
\centering
\includegraphics[width=4.0in]{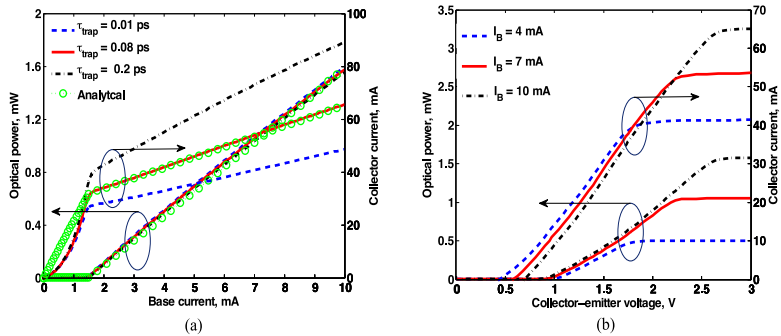}
\caption{(a) Optical power and $I_C$ as a function of $I_B$ at $V_{CE} = 4~V$ with varied $\tau_{trap}$; (b) Optical power and $I_C$ as a function of $V_{CE}$ with varied $I_B$.} \label{LIV}
\end{figure}

In the active state of transistor operation, the emitter-base junction is forward biased and the collector-base junction is reversed biased, which is evident in the band diagram (Fig. \ref{structure}). Fig. \ref{LIV} (a) shows the simulated electrical output (collector current $I_C$) and optical output as functions of the input signal (base current $I_B$). The electrical gain drops suddenly (from 21.2 to 3.9 for $\tau_{trap} =  0.08~ps$) at threshold (1.5 mA) due to the stimulated emission. We notice that while the L-I curves are almost unaffected by the trapping time, the collector current is very sensitive to the capture/escape process. This is because the stimulated emission is dominated by the hole current injected into the base, however, the electrons can either be recombined with the holes in the optical collector (i.e. QWs) or leave the transistor via the electrical collector. In the following simulation, we use $\tau_{trap} =  0.08~ps$ that corresponds to a capture time of $\sim 1~ps$ at threshold \cite{Analytical}. 
%
Due to the three-port operation and device geometry, the optical power of a T-VCSEL is controlled both  by the base current (which is determined by the base recombinations) and the collector-emitter voltage $V_{CE}$ (that controls the effective current injected into the optical cavity) \cite{TXVCSEL}. As shown in Fig. \ref{LIV} (b), the optical output follows the same trend as the collector current, which means that we can potentially monitor the optical performance by the electrical output. This voltage-controlled operation may find applications in optical communications and signal processing \cite{TunnelTL_mixing}.
\begin{figure}
\centering
\includegraphics[width=3.35in]{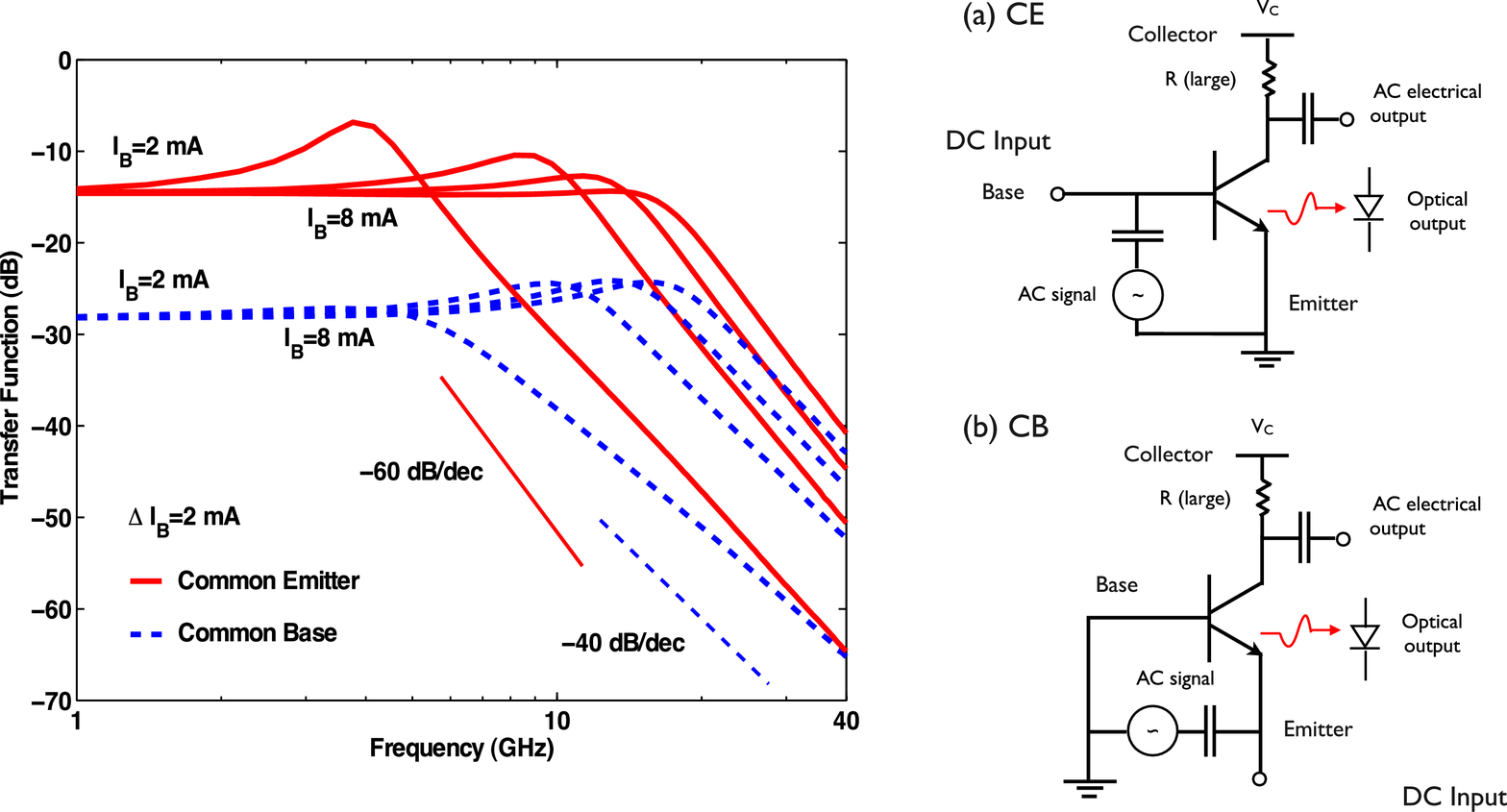}
\caption{Frequency responses of the small-signal modulation of the T-VCSEL in the CE and CB configurations. The insets show (a) the CE configuration and (b) the CB configuration.} \label{s21}
\end{figure}
\section{Modulation response}
\begin{figure}
\centering
(a)\includegraphics[width=1.32in]{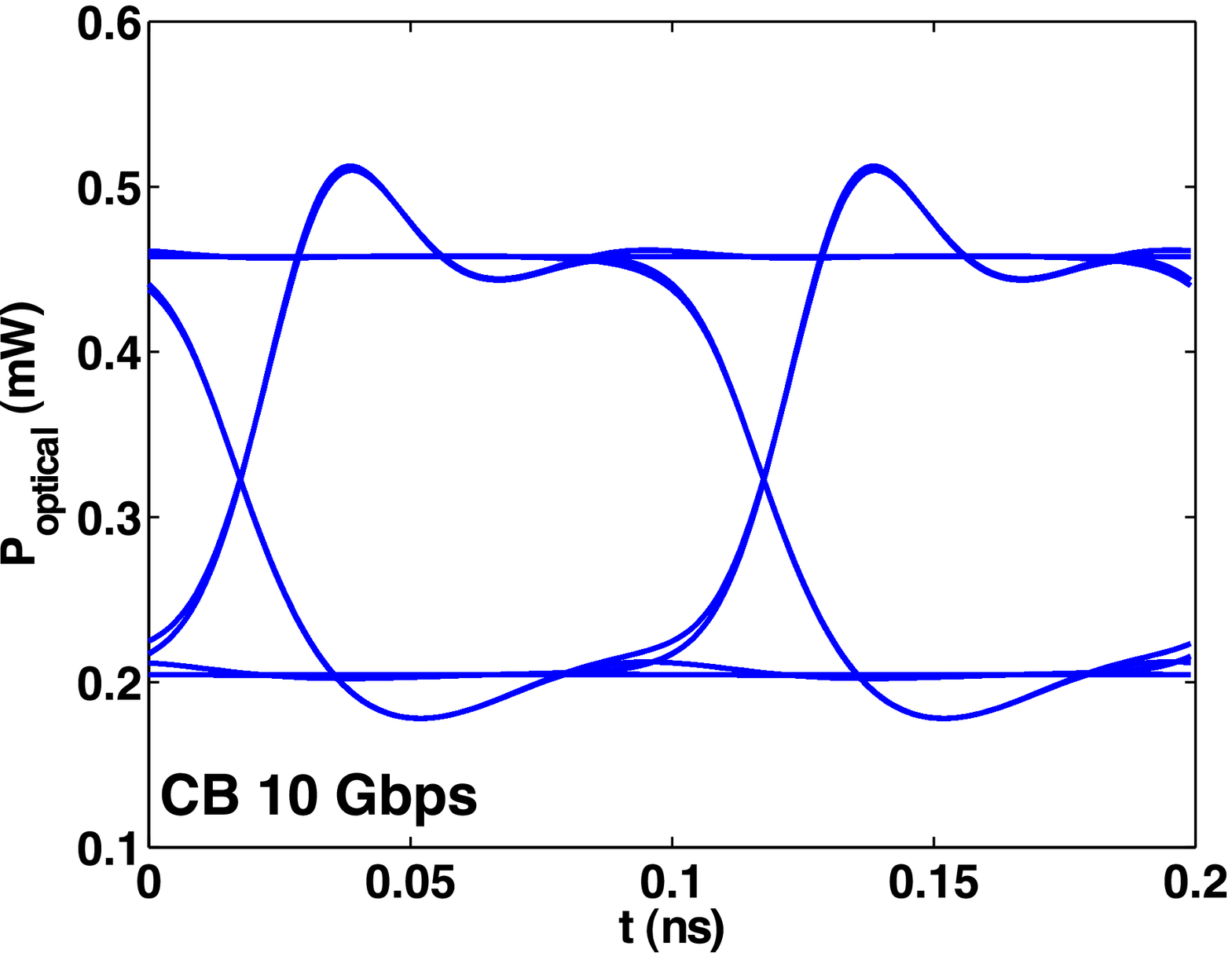}
(b)\includegraphics[width=1.32in]{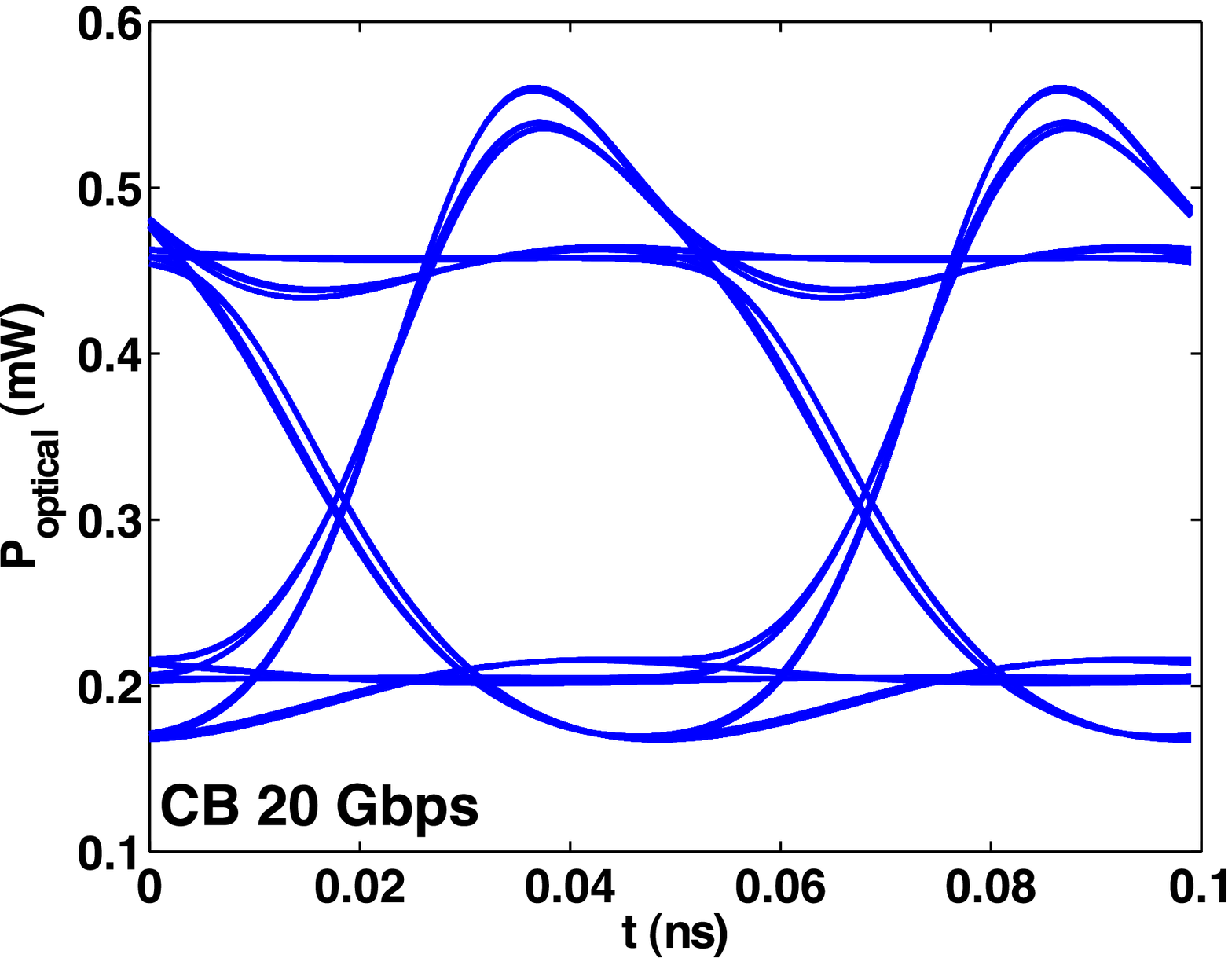}
(c)\includegraphics[width=1.32in]{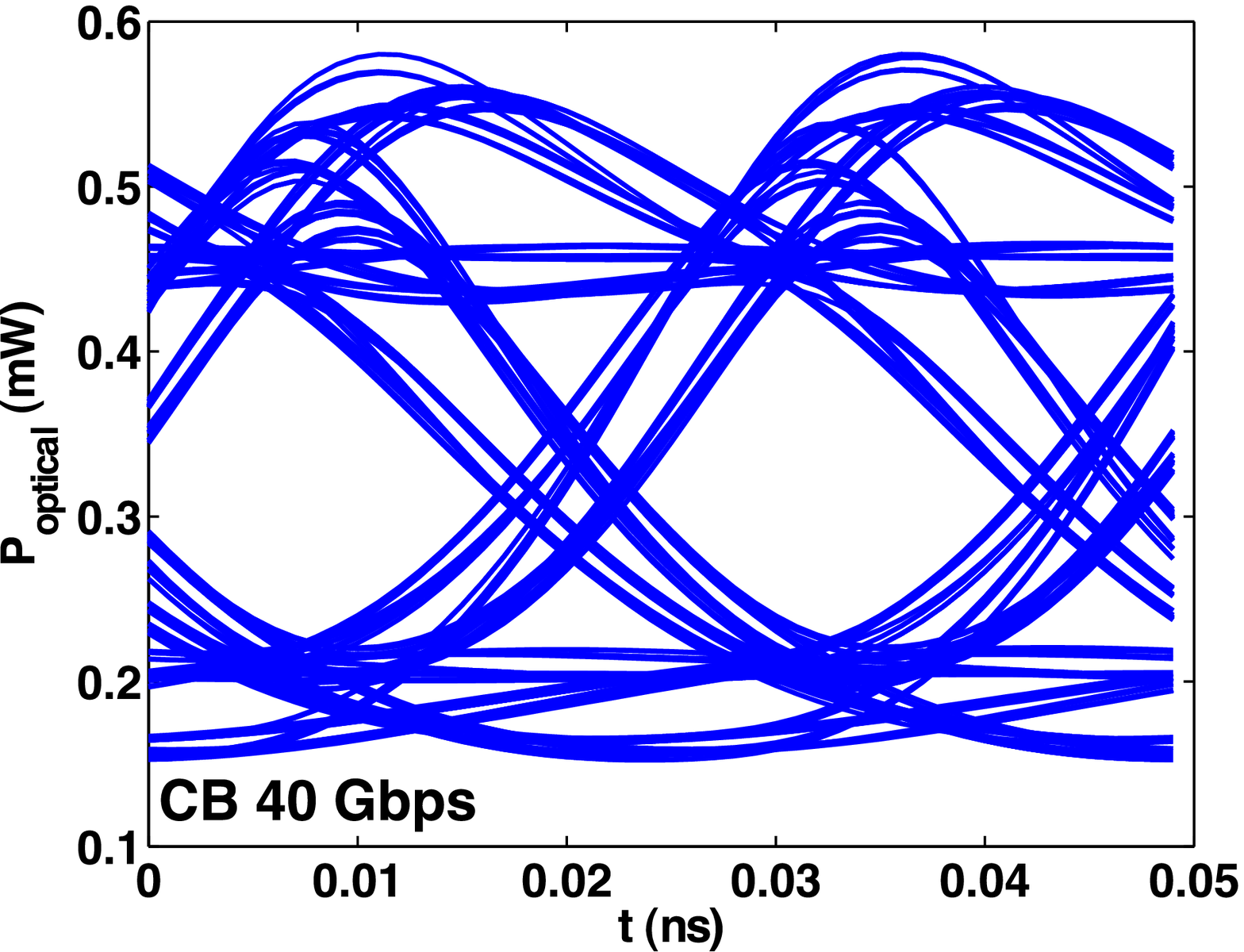}
(d)\includegraphics[width=1.32in]{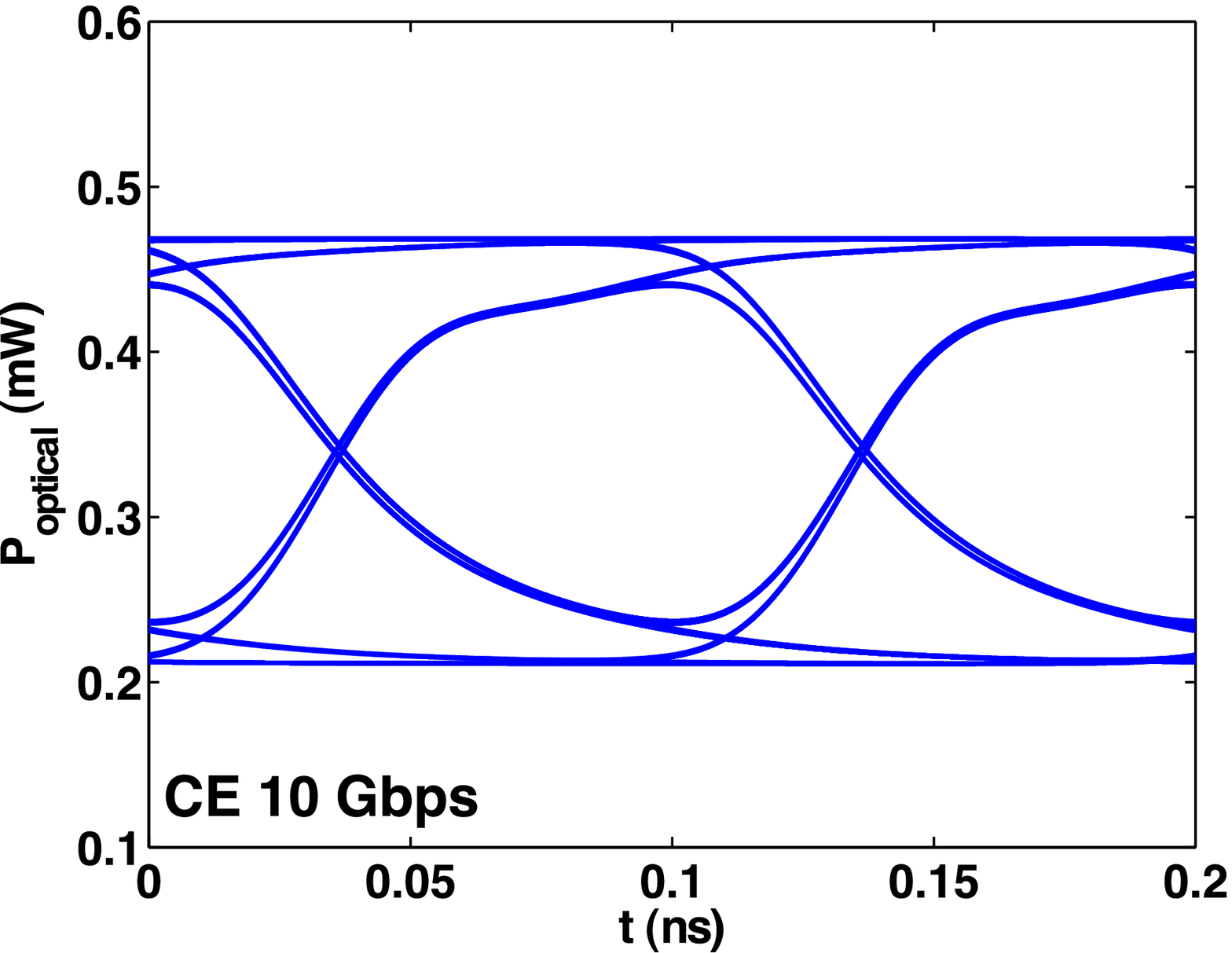}
(e)\includegraphics[width=1.32in]{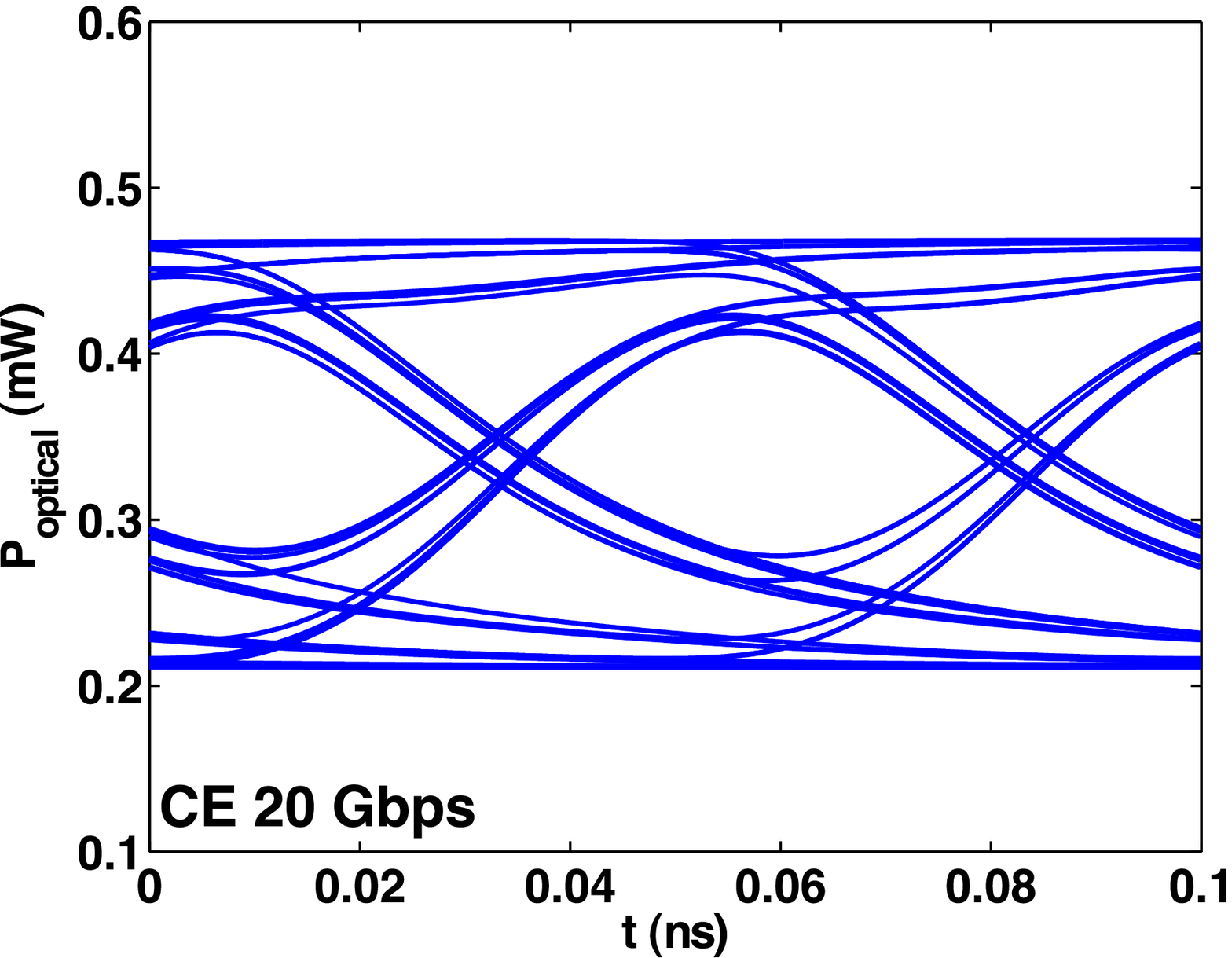}
(f)\includegraphics[width=1.32in]{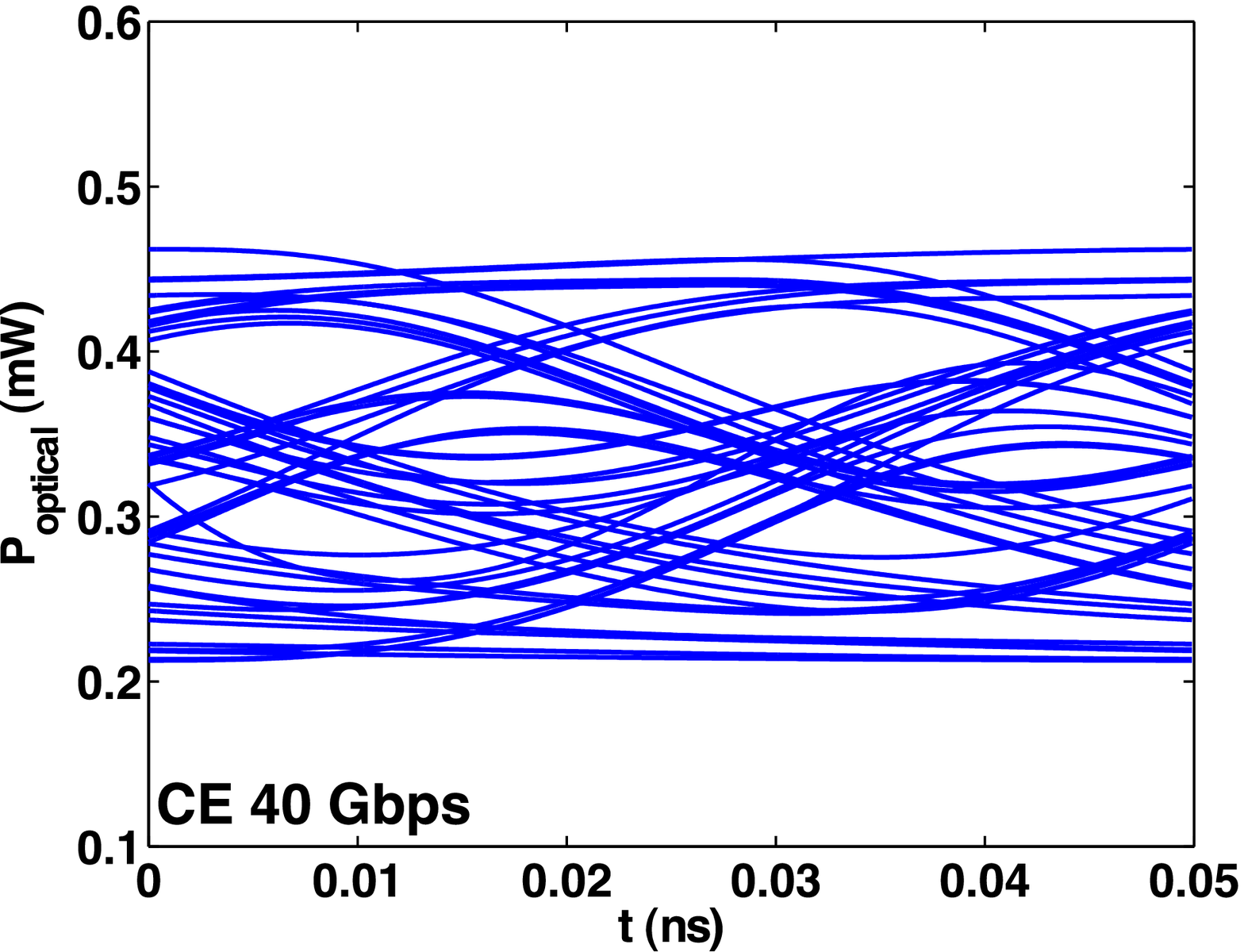}
\caption{\EDIT{Eye-diagrams of the digital modulation of the T-VCSEL: (a), (b), and (c) are for the CB configuration at a bit-rate of 10, 20, and 40 Gbps, respectively, where  $I_{B}$ is varied from 2.6 mA to 5 mA; (d), (e), and (f) are for the CE configuration at a bit-rate of 10, 20, and 40 Gbps, respectively, where $I_{E}$ is varied from 42 mA to 53 mA. The modulation currents are chosen so that the output optical powers are the same for both configurations. }} \label{dm}
\end{figure}

Several analytic models, including the charge-control model \cite{Microwave} and the carrier-transport model that considers the quantum capture/escape process \cite{Analytical} have been developed.  These models have been used to investigate the transistor lasers' frequency response \cite{Microwave,Analytical} and large-signal modulation \cite{Zhang2009,Mizuki_CLEO2010}. Here we use the model developed in \cite{Analytical} and solve the equations numerically by the finite difference method \cite{taflove}. Using the parameters extracted from the numerical modeling by PICS3D, the analytic model matches well the DC results from the numerical simulation, as shown in Fig. \ref{LIV} (a). Knowing the minority carrier distribution, we can study the T-VCSEL's modulation behavior under different configurations, i.e., common-emitter (CE) or common-base (CB). Fig.~\ref{s21} shows the frequency responses of the T-VCSEL in both configurations. We can see that while the CE configuration has the same response as normal laser diodes (as shown in \cite{Analytical}), the CB configuration demonstrates a -40 dB/dec decay after relaxation oscillation and a bandwidth enhancement, albeit, a reduction of the DC response due to the transistor current gain \cite{Analytical}. For the large-signal modulation, we report the eye-diagrams for both configurations. Figs. \ref{dm} (a)-(c) show the eye-diagrams of the T-VCSEL, in the CB configuration, directly modulated at 10, 20, and 40 Gbps, respectively. Figs. \ref{dm} (e)-(f) are for the CE configuration.  
We can see that while the ``eye" of the CE configuration starts closing at 20 Gbps, the CB configuration has an open eye-diagram up to 40 Gbps due to a larger bandwidth. 
\section{Conclusions}
We have demonstrated the design and modeling of an InGaAs MQW T-VCSEL, a vertical-cavity laser with an extra electrical output coupled to its optical output. The model has been verified by modeling a conventional VCSEL and matching the simulation results with experiments. It has been shown that the quantum capture/escape process (described by a quantum-trap model in this work) in the base of the T-VCSEL significantly affects its electrical output and, although less important for its optical DC performance, needs to be carefully treated for calculating its optical frequency response as it is highly related to the transistor's electrical gain. Our simulation has predicted a bandwidth enhancement and better large-signal performance of the T-VCSEL in the CB configuration. With a compact size, large-scale-integration capacity, a high bandwidth, and flexible three-port operation, the T-VCSEL should find novel applications in optical communications, optoelectronic data processing, and optical interconnects.
%
%
%
%
%

%
\bibliographystyle{spiebib}   
\bibliography{TVCSEL}   
%
\end{document}